\documentstyle[12pt]{article}

\begin{document}
\thispagestyle{empty}

\begin{center}
\LARGE \tt \bf{On non-Riemannian geometry of superfluids}
\end{center}

\vspace{1cm}

\begin{center} {\large L.C. Garcia de Andrade\footnote{Departamento de
F\'{\i}sica Te\'{o}rica - Instituto de F\'{\i}sica - UERJ

Rua S\~{a}o Fco. Xavier 524, Rio de Janeiro, RJ

Maracan\~{a}, CEP:20550-003 , Brasil.

E-Mail.: garcia@dft.if.uerj.br}}
\end{center}

\vspace{1.0cm}

\begin{abstract}
The Gross-Pitaevski (GP) equation describing helium superfluids is extended to non-Riemannian spacetime background where torsion is shown to induce the splitting in the potential energy of the flow. A cylindrically symmetric solution for Minkowski background with constant torsion is obtained which shows that torsion induces a damping on the superfluid flow velocity. The Sagnac phase shift is computed from the superfluid flow velocity obtained from the solution of GP equations.
\end{abstract}      
\vspace{1.0cm}       
\begin{center}
\Large{PACS number(s) : 0420,0450}
\end{center}

\newpage
\pagestyle{myheadings}
\markright{\underline{non-Riemannian geometry of superfluids.}}
\paragraph*{}
\section{Introduction}
Earlier Anandan \cite{1} and Papini \cite{2} have considered the investigation of vortices in superfluids by making use of Cooper pairs by making use of Weyl-Dirac equation. In this paper we extend the Riemannian GP equation investigated by Anandan \cite{1} to the non-Riemannian spacetime background by making use of a minimal coupling proceedure to spacetime with Cartan torsion \cite{3}. This procedure is analogous to the one used by Kleinert \cite{4} in extending the Schr\"{o}dinger equation in metric-affine spaces and also produces analogous effects as the energy splitting  obtained by L\"{a}mmerzahl \cite{5} on the non-Riemannian extended version of Dirac equation for the hydrogen-like atoms. The present paper is organised as follows: In section $2$ we present the complex extended version of GP equation to the Riemann-Cartan (RC) background where we show that the separation of the complex equation into two real equations produces an equation which is the flow conservation equation in RC spacetime while the other gives us the fluctuation in the order parameter. In section $3$ we obtain a particular solution for the $3 D$ GP equation with cylindrical symmetry. It is shown that torsion effect induces a damping into the flow superfluid velocity. The Sagnac phase shift is also computed in this section. Comments and future prospects are discussed in section $4$.
\section{The GP equation in RC spacetime}
Let us consider the time-dependent GP equation of helium superfluids in flat Minkowski $M_{4}$ space-time given by
\begin{equation} 
i{\hbar}\frac{{\partial}{\psi}}{{\partial}t}= -(\frac{{\hbar}^{2}}{2m}){{\nabla}^{2}}{\psi}+g{|{\psi}|}^{2}{\psi}
\label{1}
\end{equation}
which general relativistic generalization \cite{1} is 
\begin{equation}
{\Box}{\psi}+\frac{m^{2}c^{2}}{{\hbar}^{2}}{\psi}= -\frac{2mg}{{\hbar}^{2}}|{\psi}^{2}|{\psi}
\label{2}
\end{equation}
where ${\Box}=g^{{\mu}{\nu}}{\nabla}_{\mu}{\nabla}_{\nu}$ is the Laplace-Beltrami wave equation operator and ${\psi}$ is the order-parameter field of the quantum liquid. Here the operator ${\nabla}_{\mu}$ is the Riemannian covariant derivative, $g^{{\mu}{\nu}}$ is the contravariant metric components of the background space-time. Latin indices run over space coordinates while greek letters take into account of four-dimensional spacetime indices. The generalization to RC spacetime can be obtained by taking into account the minimal coupling principle which takes Riemannian covariant derivatives to non-Riemannian one in the sense that the Laplace-Beltrami operator applied to the field ${\psi}$ becomes 
\begin{equation}
{\Box}^{RC}{\psi}= {\Box}{\psi}- {{K^{\mu}}_{\mu}}^{\alpha}{\nabla}_{\alpha}{\psi}
\label{3}
\end{equation}
where $ {K}_{{\mu}{\nu}{\alpha}} $ is the contortion $(defect)$ tensor given by
\begin{equation}
{K^{\mu}}_{{\nu}{\alpha}}={{{\Gamma}^{RC}}^{\mu}}_{{\nu}{\alpha}}- {{{\Gamma}}^{\mu}}_{{\nu}{\alpha}}
\label{4}
\end{equation}
where $ {\Gamma} $ is the Riemannian connection. Due to the minimal coupling the complex GP equation becomes 
\begin{equation}
{\Box}{\psi}+\frac{m^{2}c^{2}}{{\hbar}^{2}}{\psi}={{K^{\mu}}_{\mu}}^{\alpha}{\nabla}_{\alpha}{\psi} -\frac{2mg}{{\hbar}^{2}}|{\psi}^{2}|{\psi}
\label{5}
\end{equation}
To solve this complex equation we split the complex equation into two real equations with the help of the complex wave function ${\psi}={\alpha}e^{i{\phi}}$. Here the amplitude ${\alpha}$ and the phase shift ${\phi}$ are real functions. This procedure yields two real equations 
\begin{equation}
g_{{\mu}{\nu}}v^{\mu}v^{\nu}= 1+f({\alpha})-\frac{{\hbar}^{2}}{m^{2}c^{2}}{{K^{\mu}}_{\mu}}^{\rho}{{\partial}_{\rho}ln{\alpha}}
\label{6}
\end{equation}
where the function $f({\alpha})$ is given by
\begin{equation} 
f({\alpha}):= \frac{{\hbar}^{2}}{m^{2}c^{2}}\frac{{\Box{\alpha}}}{\alpha}+\frac{2g}{m^{2}c^{2}}{\alpha}^{2}
\label{7}
\end{equation}
Here 
\begin{equation}
v_{\mu}= -\frac{\hbar}{mc}{\partial}_{\mu}{\phi}
\label{8}
\end{equation}
is the superfluid flow velocity. The other real equation is nothing less than the matter current conservation in RC manifold
\begin{equation}
{\nabla}_{\mu}J^{\mu}= {{K^{\mu}}_{\mu}}^{\rho} J_{\rho}
\label{9}
\end{equation}
where to simplify matters we have considered $ J^{\mu}={\alpha}^{2}v^{\mu}$ which is the flow superfluid current. 
\section{The GP equation in Minkowski spacetime with torsion}
Since we are interested especially on the torsion effects on the superfluid flow we consider in this section that our manifold is composed of the Minkowski space-time endowed with torsion. Let us now substitute expression (\ref{8}) into (\ref{6}) leads to
\begin{equation}
[-\frac{{\hbar}^{2}}{m^{2}c^{2}}({\nabla}^{2}-({\nabla}{\phi})^{2})-(1+\frac{2g{\alpha}^{2}}{mc^{2}}-\vec{K}.{\nabla})]{\alpha}=0
\label{10}
\end{equation}
where we have consider the manifold to be reduced to a $3-D$ space while the other conservation current reduces to 
\begin{equation}
{\partial}_{i}[{\alpha}^{2}{\partial}^{i}{\phi}]=-{\alpha}^{2}K^{i}{\partial}_{i}{\phi}
\label{11}
\end{equation}
In these last two expressions we have consider the effect of torsion on the $3-D$ GP equations. Here we note that $K^{i}=\vec{K}= {{K^{\mu}}_{\mu}}^{i}$. From expression (\ref{10}) we note that the term $ \vec{K}.{\nabla}{\alpha}$ represents the energy splitting torsion effect on the potential energy of the quantum liquid. An analogous result have been considered by L\"{a}mmerzahl \cite{5} on the energy splitting of Dirac equation in an attempt to use the Hughes-Drever effect to detect torsion effects on hydrogen-like atoms. The real form of the GP equations represent a system of two coupled differential equations , thus we have to solve one of the equations to substitute into the other to complete the solution of the PDE. Let us now solve firsthe conservation current equation (\ref{12}) which reduces to 
\begin{equation}
{\nabla}^{2}{\phi}+[K_{i}+\frac{{\partial}_{i}{\alpha}^{2}}{{\alpha}^{2}}]{\partial}^{i}{\phi}=0
\label{12}
\end{equation}
Taking into account that the ${\nabla}{\phi}$ is proportional to the superfluid velocity $\vec{v}$ we can express equation (\ref{12}) in terms of the flow velocity as
\begin{equation}
div{\vec{v}}+\vec{K}.\vec{v}=0
\label{13}
\end{equation}
where we have considered the following approximation $\frac{{\partial}_{i}{\alpha}^{2}}{{\alpha}^{2}}=0$.  By considering a cylindrically \cite{6} symmetric solution of torsion vortex type, we obtain 
\begin{equation}
{\nabla}^{2}{\phi}+K^{r}{\partial}_{r}{\phi}=0
\label{14}
\end{equation}
or
\begin{equation}
{\partial}_{r}v_{r}+({K+\frac{1}{r}})v_{r}=0
\label{15}
\end{equation}
where , to simplify matters ,we have consider a constant background torsion.
The solution of this equation yields
\begin{equation}
v_{r}= \frac{e^{-Kr}}{r}
\label{16}
\end{equation}
which physically shows that torsion induces a damping effect on the superfluid flow velocity. In cosmology torsion also induces damping effects \cite{3}. With the flow velocity in hands we are now able to compute the Sagnac phase shift before going to solve the differential equation for the amplitude ${alpha}$. Let us now compute the Sagnac phase shift as
\begin{equation}
{\Delta}{\phi}=\frac{mc}{\hbar}{\oint}_{\gamma}v_{\mu}dx^{\mu}
\label{17}
\end{equation}
Substitution of (\ref{16}) into (\ref{17}) reads
\begin{equation}
{\Delta}{\phi}= \frac{mc}{\hbar}[lnr-Kr]
\label{18}
\end{equation}
Let us now solve the GP amplitude equation. In cylindrical coordinates this equation reduces to
\begin{equation}
[{{\partial}_{r}}^{2}+\frac{(1+2\frac{m^{2}c^{2}K}{{\hbar}^{2}})}{r}{\partial}_{r}-\frac{\frac{m^{2}c^{2}}{{\hbar}^{2}}}{r^{2}}]{\alpha}=0
\label{19}
\end{equation}
where we have used the approximation close to the torsion vortex axis $r<<<1$. Equation (\ref{19}) is a Bessel type equation and therefore we shall use the ansatz ${\alpha}= A_{n}J_{n}=A_{n}r^{n}$. Substitution of this ansatz into equation (\ref{19}) one obtains the auxialiry algebraic equation
\begin{equation}
n^{2}+{2\frac{m^{2}c^{2}K}{{\hbar}^{2}}}n-\frac{m^{2}c^{2}}{{\hbar}^{2}}=0
\label{20}
\end{equation}
Solutions of this equations are
\begin{equation}
n = -{\frac{m^{2}c^{2}K}{{\hbar}^{2}}}(1{\mp}\sqrt{1-\frac{{\hbar}^{2}}{m^{2}c^{2}K^{2}}}
\label{21}
\end{equation}
Therefore the complete solution is given by ${\psi}= A_{n}J_{n}e^{i{\phi}}$ with the values of n and ${\phi}$ given respectively by (\ref{21}) and  
(\ref{18}) above.
\section{Conclusions}
The ideas discussed and presented here maybe useful also in astrophysical problems such as the investigation of the interior of neutron stars with superfluid cores. We could imagine that the inner core of the star could be the source of torsion which could affect the behaviour of more external superfluid layers. Also recently analog models of general relativity \cite{7} using scattering of phonons in superfluids could be investigating by using this model or translational holonomy techniques \cite{8}.

\section*{Acknowledgments}

\paragraph*{}
Thanks are due to  J. Anandan,M. Visser and Yu N. Obukhov for helpful discussions on the subject of this paper. I am very much indebt to Universidade do Estado do Rio de Janeiro(UERJ) and CNPq. (Brazilian Government Agency) for financial support.

\end{document}